\documentclass[aps,prb,showpacs,twocolumn,superscriptaddress]{revtex4-1}
\usepackage{graphicx,amsmath,amssymb}
\usepackage[usenames]{color}
\usepackage{indentfirst}
\usepackage{float}
\usepackage[T1]{fontenc}
\usepackage[colorlinks=true, citecolor=blue, urlcolor=blue, linkcolor=blue ]{hyperref}
\hypersetup{breaklinks=true}
\begin{document}

\bibliographystyle{apsrev4-1}

\title{Lifshitz Transition in Triangular Lattice Kondo-Heisenberg Model}
\author{Lan Zhang}
\affiliation{School of Physical Science and Technology $\&$ Key Laboratory for
Magnetism and Magnetic Materials of the MoE, Lanzhou University, Lanzhou 730000, China}
\author{Hong-Gang Luo}
\email{luohg@lzu.edu.cn}
\affiliation{School of Physical Science and Technology $\&$ Key Laboratory for
Magnetism and Magnetic Materials of the MoE, Lanzhou University, Lanzhou 730000, China}
\affiliation{Beijing Computational Science Research Center, Beijing 100084, China}
\author{Yin Zhong}
\email{zhongy@lzu.edu.cn}
\affiliation{School of Physical Science and Technology $\&$ Key Laboratory for
Magnetism and Magnetic Materials of the MoE, Lanzhou University, Lanzhou 730000, China}
\begin{abstract}
Motivated by recent experimental progress on triangular lattice heavy-fermion compounds, we investigate possible Lifshitz transitions and the scanning tunnel microscope (STM) spectra of the Kondo-Heisenberg model on the triangular lattice. In the heavy Fermi liquid state, the introduced Heisenberg antiferromagnetic interaction ($J_H$) results in the twice Lifshitz transition at the case of the nearest-neighbour electron hopping but with next-nearest-neighbour hole hopping and the case of the nearest-neighbour hole hopping but with next-nearest-neighbour electron hopping, respectively. Driven by $J_H$, the Lifshitz transitions on triangular lattice are all continuous in contrast to the case in square lattice. Furthermore, the STM spectra shows rich line-shape which is influenced by the Kondo coupling $J_{K}$, $J_{H}$ and the ratio of the tunneling amplitude $t_{f}$ versus $t_{c}$. Our work provides a possible scenario to understand the Fermi surface topology and the quantum critical point in heavy-fermion compounds.
\end{abstract}

\maketitle
\section{Introduction}
The Lifshitz transition, where the Fermi surface (FS) topology changes,\cite{lifshitz1960} is beyond the paradigm of Landau's symmetry breaking theory.
This unconventional transition has been observed experimentally in cuprate superconductors,\cite{PhysRevB.81.180513,PhysRevLett.114.147001,PhysRevB.83.054506} iron-based superconductors,\cite{PhysRevLett.112.156401,PhysRevB.89.224517,cho2016energy,PhysRevB.90.224508,PhysRevLett.103.047002,PhysRevB.83.020501,PhysRevB.86.165117,
PhysRevB.88.220508,liu2010evidence} topological insulator,\cite{volovik2017topological} graphene\cite{PhysRevB.96.155432} and heavy-fermion compounds.\cite{PhysRevLett.96.026401,PhysRevLett.99.056401,PhysRevB.79.214428,PhysRevB.83.115133,PhysRevLett.110.256403,PhysRevLett.116.037202}
Particularly, for some quantum critical heavy-fermion materials, such as YbRh$_{2}$Si$_{2}$, its magnetic field dependent thermopower, thermal conductivity, resistivity and Hall effect shows three transitions at high fields and the Lifshitz transitions are argued to be their origin.\cite{PhysRevLett.110.256403} For CeRu$_{2}$Si$_{2}$, the high resolution Hall effect and magnetoresistance measurements across the metamagnetic transition are explained as an abrupt $f$-electron localization, where one of the spin-split sheets of the heaviest Fermi surface shrink to a point.\cite{PhysRevLett.96.026401} The Lifshitz transition leads to the way to understand the relation of the FS topology and the quantum critical point in heavy-fermion systems.\cite{paschen2004}

Theoretically, the Lifshitz transition in heavy fermion systems have been carefully explored with mean-field theory and dynamical mean-field theory.
\cite{PhysRevB.83.033102,zhong2015fermionology,LIU2014,PhysRevB.86.075108,PhysRevB.94.155103,PhysRevLett.110.226403,PhysRevB.98.045105,PhysRevLett.45.1028,
PhysRevB.61.3435,PhysRevLett.110.026403,PhysRevLett.111.026401,PhysRevB.87.205144} At the mean-field level, the Lifshitz transition is triggered with the introduction of Heisenberg coupling into the usual Kondo lattice model, i.e. Kondo-Heisenberg model (KHM), and a case studying on square lattice suggests both first and second-order Lifshitz transitions.\cite{PhysRevB.83.033102,zhong2015fermionology,LIU2014} Interestingly, the appearance of Lifshitz transition with enhanced antiferromagnetic Heisenberg interaction preempts the disentanglement of Kondo singlet, thus the resulting Kondo breakdown mechanism predicted in literature should be reexamined.\cite{PhysRevB.86.075108,PhysRevB.77.134439,PhysRevLett.106.137002}

Recently, non-Fermi liquid behaviors have been observed in triangular lattice heavy-fermion compounds like YbAgGe and YbAl$_{3}$C$_{3}$.\cite{PhysRevB.71.054408,PhysRevLett.110.176402,PhysRevB.69.014415,PhysRevLett.111.116401,PhysRevB.87.220406,PhysRevB.85.144416} Due to the frustration effect introduced by local $f$-electron spin located on the triangular lattice, the observed non-Fermi liquid phenomena could be linked to the idea of Kondo breakdown, where critical Kondo boson and deconfined gauge field induce singularity in thermodynamics and transport.\cite{PhysRevB.71.054408,PhysRevB.69.014415,PhysRevB.85.144416} However, as exemplified by the study on the square lattice, the topology of FS may change radically before any noticeable breakdown of Kondo effect, therefore the possibility of Lifshitz transition on triangular lattice should be investigated firstly.

In the present work, we employ the large-$N$ mean-field approach to study the KHM on the triangular lattice. As expected, we find that the Heisenberg antiferromagnetic interaction ($J_H$) induces twice FS topology change at the case of the nearest-neighbour (NN) electron hopping but with next-nearest-neighbour (NNN) hole hopping and the case of the NN hole hopping but with NNN electron hopping. Both Lifshitz transitions are continuous, which is different from the square lattice case, i.e. the first-order and the second-order phase transition.\cite{PhysRevB.83.033102} The density of state (DOS) of conduction electron is changed by $J_H$. To meet with experiments, we give the STM line-shape of the differential conductance $dI/dV$ for different Kondo coupling ($J_{K}$) and the ratio of the tunneling amplitude of $f$-electron $t_{f}$ versus conduction electron's $t_{c}$. The calculated spectra are qualitatively consistent with data in CeCoIn$_{5}$.\cite{aynajian2012visualizing}

The paper is organized as follows. In Sec. 2, we describe the Kondo-Heisenberg model under the large-N mean-field theory. In Sec. 3, we present the Lifshitz transition and the DOS of conduction electron. In Sec. 4, we give the line-shape of differential conductance at different Heisenberg antiferromagnetic interaction, the Kondo coupling and the ratio of the tunneling amplitude of f-electron to conduction electron's. Finally, Sec. 5 is devoted to a brief conclusion and perspective.

\section{Model and mean-field approach}\label{sec2}
The model Hamiltonian of the KHM is given by
\begin{eqnarray}
H&&=-t\sum_{<ij>\sigma}c^{\dag}_{i\sigma}c_{j\sigma}+t_{1}\sum_{\ll ij\gg\sigma}c^{\dag}_{i\sigma}c_{j\sigma}-\mu\sum_{i\sigma}c^{\dag}_{i\sigma}c_{i\sigma}\nonumber\\
 &&\quad+J_{K}\sum_{i}\textbf{S}_{i}\cdot\textbf{s}_{i}+J_{H}\sum_{<ij>}\textbf{S}_{i}\cdot\textbf{S}_{j},\label{eq1}
\end{eqnarray}
where $c^{\dag}_{i\sigma}(c_{i\sigma})$ denotes the creation (annihilation) operator of conduction electron with spin $\sigma=\uparrow,\downarrow$. The first line in Eq.~(\ref{eq1}) describes the hoppings of conduction electron and $\mu$ is the chemical potential. $\langle \cdot \rangle$ and $\langle\langle \cdot \rangle\rangle$ represent the NN and the NNN hopping, respectively. (The NNN hopping is introduced to avoid the occasional nesting.) The $J_{K}$ term in the second line denotes the Kondo coupling between the localized $f$-electron and conduction electron. $\textbf{S}_{i}= \frac{1}{2}\sum_{\alpha\beta}f_{i\alpha}^{\dag}\tau_{\alpha\beta}f_{i\beta}$ is the fermionic representation of localized $f$-electron spin with the local constraint $\sum_{\sigma} f^{\dag}_{i\sigma}f_{i\sigma}=1$, while $\textbf{s}_{i}=\frac{1}{2}\sum_{\sigma\sigma^{\prime}}c^{\dag}_{i,\sigma} \tau_{\sigma\sigma^{\prime}} c_{i,\sigma^{\prime}}$ is for conduction electron. The last $J_H$ term is the Heisenberg exchange interaction firstly introduced by Coleman and Andrei.\cite{Coleman1989} It has also been used by Iglesias \textit{et al.} to consider the antiferromagnetic long-range order in some Ce-based heavy fermion compounds.\cite{Arispe1995255,Iglesias1996160,Lacroix1997503}

To proceed, we use the fermionic large-$N$ mean-field method,\cite{Read1983} which is believed to capture qualitative features in heavy Fermi liquid states. Introducing valence-bond order parameter $\chi_{ij}=-\left\langle\Sigma_{\alpha}f^{\dag}_{i\alpha}f_{j\alpha}\right\rangle$ and Kondo hybridization parameter $V=\left\langle\Sigma_{\alpha}f^{\dag}_{i,\alpha}c_{i\alpha}\right\rangle$,\cite{PhysRevB.83.033102} and considering the uniform resonance-valence-bond ansatz $\chi_{ij}=\chi$ in Refs.~\cite{PhysRevB.83.033102,Liu2012}, one can obtain $\textbf{S}_{i}\cdot\textbf{S}_{j} =\frac{1}{2}\left[\chi(f^{\dag}_{i\uparrow}f_{j\uparrow}
+f^{\dag}_{i\downarrow}f_{j\downarrow})+H.c.\right]+\frac{\chi^2}{2}$. Based on these mean-field formulations, Eq.(\ref{eq1}) can be rewritten in the $k$-space as follows
\begin{eqnarray}
&&H=\sum_{k}\Psi^{\dag}_{k}\left(
                           \begin{array}{ccc}
                           \varepsilon_{k}-\mu & -\frac{J_{K}V}{2}\\
                           \\
                            -\frac{J_{K}V}{2} & \chi_{k}\\
                           \end{array}
                          \right)\Psi_{k}+\sum_{k\sigma}\chi_{k}f^{\dag}_{k\sigma}f_{k\sigma}\nonumber\\
  &&\qquad+N_{S}\left(\frac{J_{K}V^{2}}{2}+\frac{3J_{H}\chi^{2}}{2}-\lambda\right),\label{eq2}
\end{eqnarray}
where $\Psi^{\dag}_{\textbf{k}}=\left(c^{\dag}_{\textbf{k}\sigma}, f^{\dag}_{\textbf{k}\sigma}\right)$ is a two-component Nambu spinor, and $\gamma_{k}=2\cos(\sqrt{3}k_{y}/2)\cos(k_{x}/2)+\cos(k_{x})$, $\chi_{k}=J_{H}\chi\gamma_{k}+\lambda$ is the kinetic energy of the $f$-electron, and  $\varepsilon_{k}=2t_{1}[2\cos(\sqrt{3}k_{y}/2)\cos(3k_{x}/2)+\cos(\sqrt{3}k_{y})]-2t\gamma_{k}$ denoting the energy spectrum of conduction electron. Also, the Lagrangian multiplier $\lambda$ is introduced to impose the local constraint on average.
The quasiparticle excitation spectrum can be easily obtained by
\begin{equation}
E^{\pm}_{k}=\frac{(\varepsilon_{k}-\mu+\chi_{k})}{2}\pm\frac{\sqrt{(\varepsilon_{k}-\mu-\chi_{k})^{2}+J^{2}_{K}V^{2}}}{2}.\label{eq3}
\end{equation}
The ground-state energy of the KHM is
\begin{equation}
E_{g}=\frac{2}{N_{S}}\sum_{k,\pm}E^{\pm}_{k}\theta(-E^{\pm}_{k})+\left(\frac{J_{K}V^{2}}{2}+\frac{3J_{H}\chi^{2}}{2}-\lambda\right),
\end{equation}
where $\theta(x)$ is the step function. The factor $2$ comes from the spin degeneracy. Then, the MF equations for $\chi,V,\lambda$ can be derived by minimizing the ground-state energy and the chemical potential $\mu$ is determined by the conduction electron density $n_{c}$, i.e. $\frac{\partial{E_{g}}}{\partial{\chi}}=0, \frac{\partial{E_{g}}}{\partial{V}}=0, \frac{\partial{E_{g}}}{\partial{\lambda}}=0, -\frac{\partial{E_{g}}}{\partial{\mu}}=n_{c}$. One can get four self-consistent MF equations:
\begin{eqnarray}
&&-\frac{1}{N_{S}}\sum_{k,\pm}\theta(-E^{\pm}_{k})\gamma_{k}\left[1\mp\frac{(\varepsilon_{k}-\mu-\chi_{k})}{W_{k}}\right]=3\chi,\\
&&-\frac{1}{N_{S}}\sum_{k,\pm}\theta(-E^{\pm}_{k})\left[\frac{\pm J_{K}}{W_{k}}\right]=1,\\
&&\frac{1}{N_{S}}\sum_{k,\pm}\theta(-E^{\pm}_{k})\left[1\mp\frac{(\varepsilon_{k}-\mu-\chi_{k})}{W_{k}}\right]=1,\\
&&\frac{1}{N_{S}}\sum_{k,\pm}\theta(-E^{\pm}_{k})\left[1\pm\frac{(\varepsilon_{k}-\mu-\chi_{k})}{W_{k}}\right]=n_{c},
\end{eqnarray}
where $W_{k}=\sqrt{(\varepsilon_{k}-\mu-\chi_{k})^{2}+(J_{K}V)^{2}}$.

\section{Lifshitz transition}\label{sec4}
We consider the case of $J_{H}\ll J_{K}$, where the paramagnetic heavy Fermi liquid state is stable to other symmetry-breaking and exotic fractionalized states. When the Heisenberg interaction $J_{H}$ increases, the band structure of quasiparticle evolves and Lifshitz transition is expected to occur.
\begin{figure}[htb]
\centering
\includegraphics[width = 0.8\columnwidth]{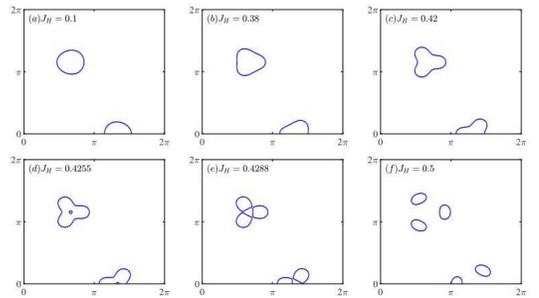}\\
\caption{\label{Fig1} The FS evolution of the KHM versus the Heisenberg interaction $J_{H}$ for NN electron hopping but with NNN hole hopping, where $t=1, t_{1}/t=0.3, n_{c}=0.9, J_{K}=2.5$. }
\end{figure}
\begin{figure}[htb]
\centering
\includegraphics[width = 0.8\columnwidth]{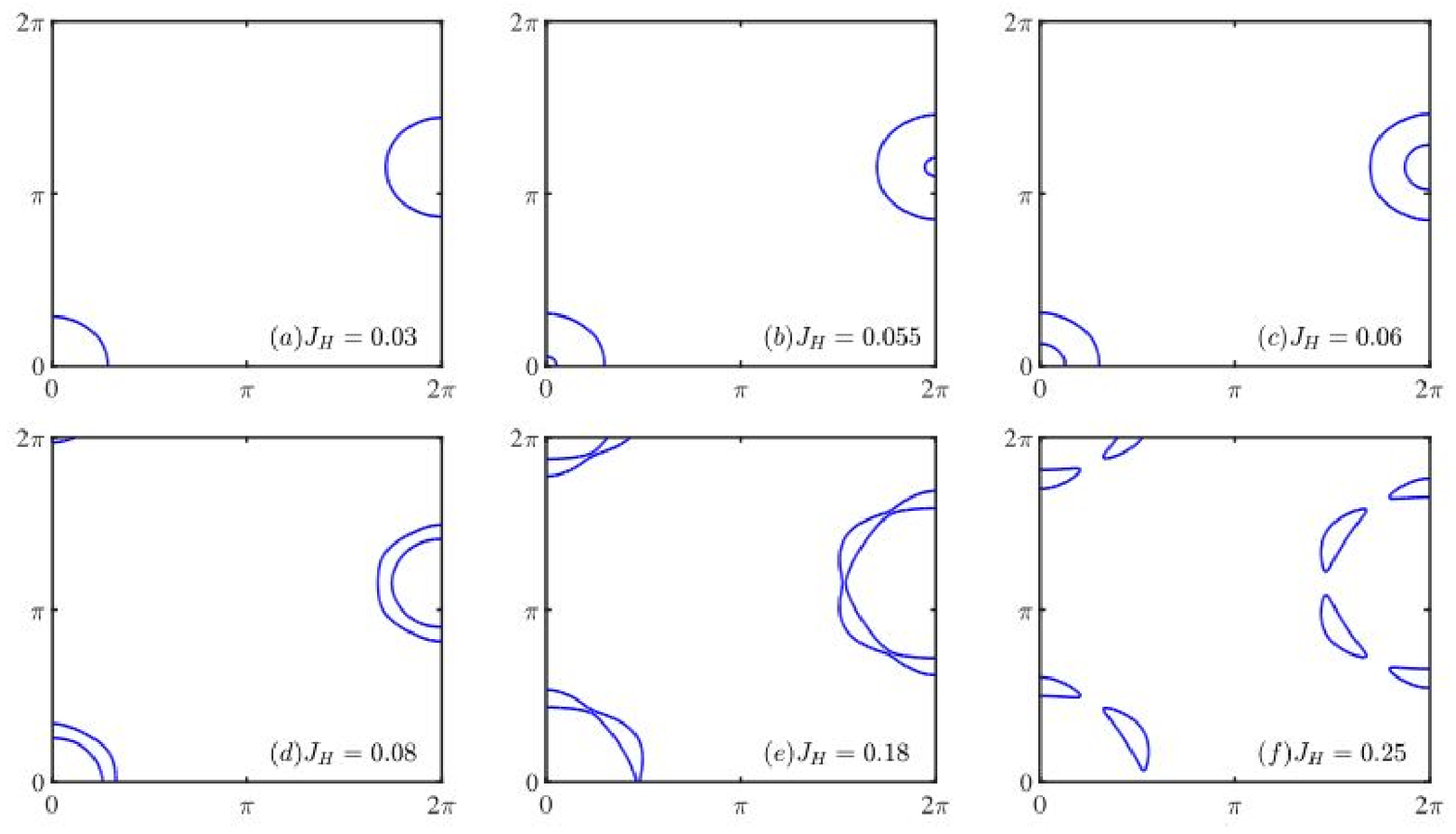}\\
\caption{\label{Fig2} The FS evolution of the KHM versus the Heisenberg interaction $J_{H}$ for NN hole hopping but with NNN electron hopping, where $t=-1, t_{1}/t=0.3, n_{c}=0.9, J_{K}=2.5$. }
\end{figure}
\begin{figure}[htb]
\centering
\includegraphics[width = 0.7\columnwidth]{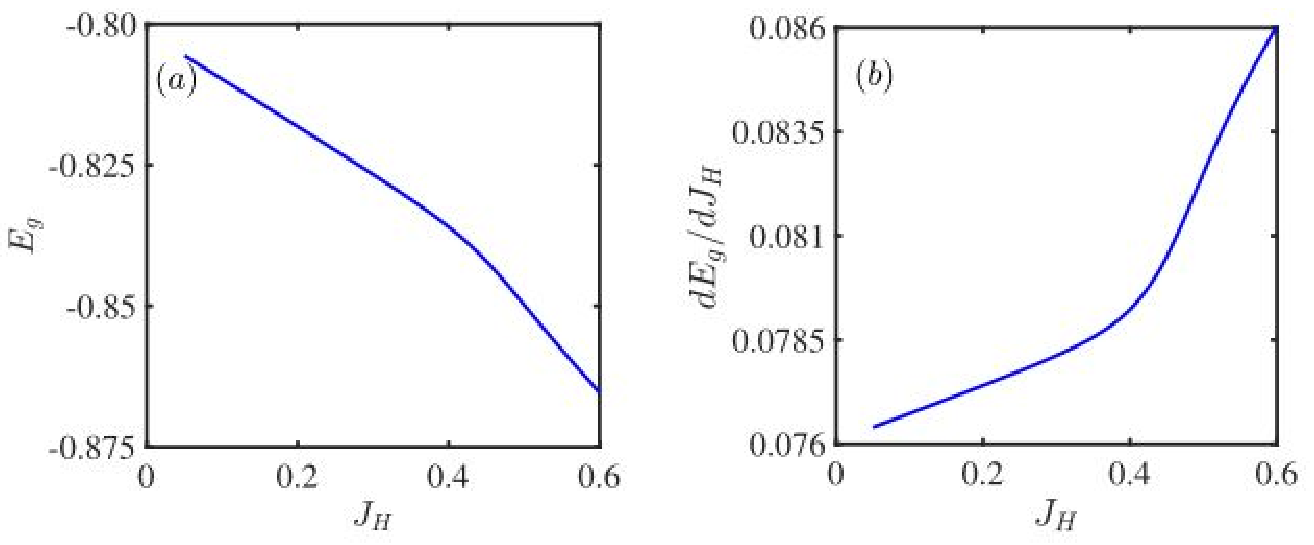}\\
\caption{\label{Fig9} The ground-state energy $E_{g}$ (subplot (a)) and the first derivative $dE_{g}/dJ_{H}$ (subplot (b)) versus the Heisenberg interaction $J_{H}$ for NN electron hopping but with NNN hole hopping, where $t=1, t_{1}/t=0.3, n_{c}=0.9, J_{K}=2.5$. }
\end{figure}
\begin{figure}[htb]
\centering
\includegraphics[width = 0.7\columnwidth]{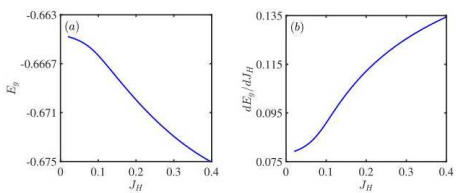}\\
\caption{\label{Fig10} The ground-state energy $E_{g}$ (subplot (a)) and the first derivative $dE_{g}/dJ_{H}$ (subplot (b)) versus the Heisenberg interaction $J_{H}$ for NN hole hopping but with NNN electron hopping, where $t=-1, t_{1}/t=0.3, n_{c}=0.9, J_{K}=2.5$. }
\end{figure}

In Fig.~\ref{Fig1} and Fig.~\ref{Fig2}, the FS is a normal circle when $J_{H}$ is small as shown in Fig.~\ref{Fig1} (a) and Fig.~\ref{Fig2} (a), which means the influence of the short-range antiferromagntic correlation is negligible. However, when $J_{H}$ is increasing, the short-range antiferromagnetic correlation starts to change the electronic structure, and the FS begins to deform. Our system have twice Lifshitz transition in two cases as shown in Fig.~\ref{Fig1} (d), (e), and Fig.~\ref{Fig2} (b), (e). In Fig.~\ref{Fig1} (d) and Fig.~\ref{Fig2} (b), there emerges a small circle below FS at the center, the particles begin to fill the area between two loops. In Fig.~\ref{Fig1} (e) and Fig.~\ref{Fig2} (e), the FS happens to split into many Fermi pockets  after this critical point, each pocket is the FS of electrons, and they will be shifted inward along the direction M $\rightarrow$ $\Gamma$ associated with $J_{H}$, such as Fig.~\ref{Fig1} (f) and Fig.~\ref{Fig2} (f). The quantum critical points for NN electron hopping with NNN hole hopping has the larger Heisenberg coupling $J_{H}$ than the case of the NN hole hopping with NNN electron hopping.

Due to many experiments on heavy-fermion quantum critical compounds YbRh$_{2}$Si$_{2}$ and CeRu$_{2}$Si$_{2}$,\cite{PhysRevLett.110.256403,PhysRevLett.96.026401} the FS change relates to the quantum phase transition. Thus, to identify the quantum phase transition around the Lifshitz transition, the ground-state energy $E_{g}$ and its first derivative $dE_{g}/dJ_{H}$ versus $J_{H}$ is shown in Fig.~\ref{Fig9} and Fig.~\ref{Fig10}, where $dE_{g}/dJ_{H}$ is given by
\begin{equation}
\frac{dE_{g}}{dJ_{H}}=\frac{1}{N_{S}}\sum_{k,\pm}\theta(-E^{\pm}_{k})\chi\gamma_{k}\left[1\mp\frac{(\varepsilon_{k}-\mu-\chi_{k})}{W_{k}}\right]+\frac{3\chi^{2}}{2}.
\end{equation}
Both lines are smooth across the changes of FS topology, which demonstrates that the Lifshitz transitions are second-order transitions.
\begin{figure}[htb]
\centering
\includegraphics[width = 0.8\columnwidth]{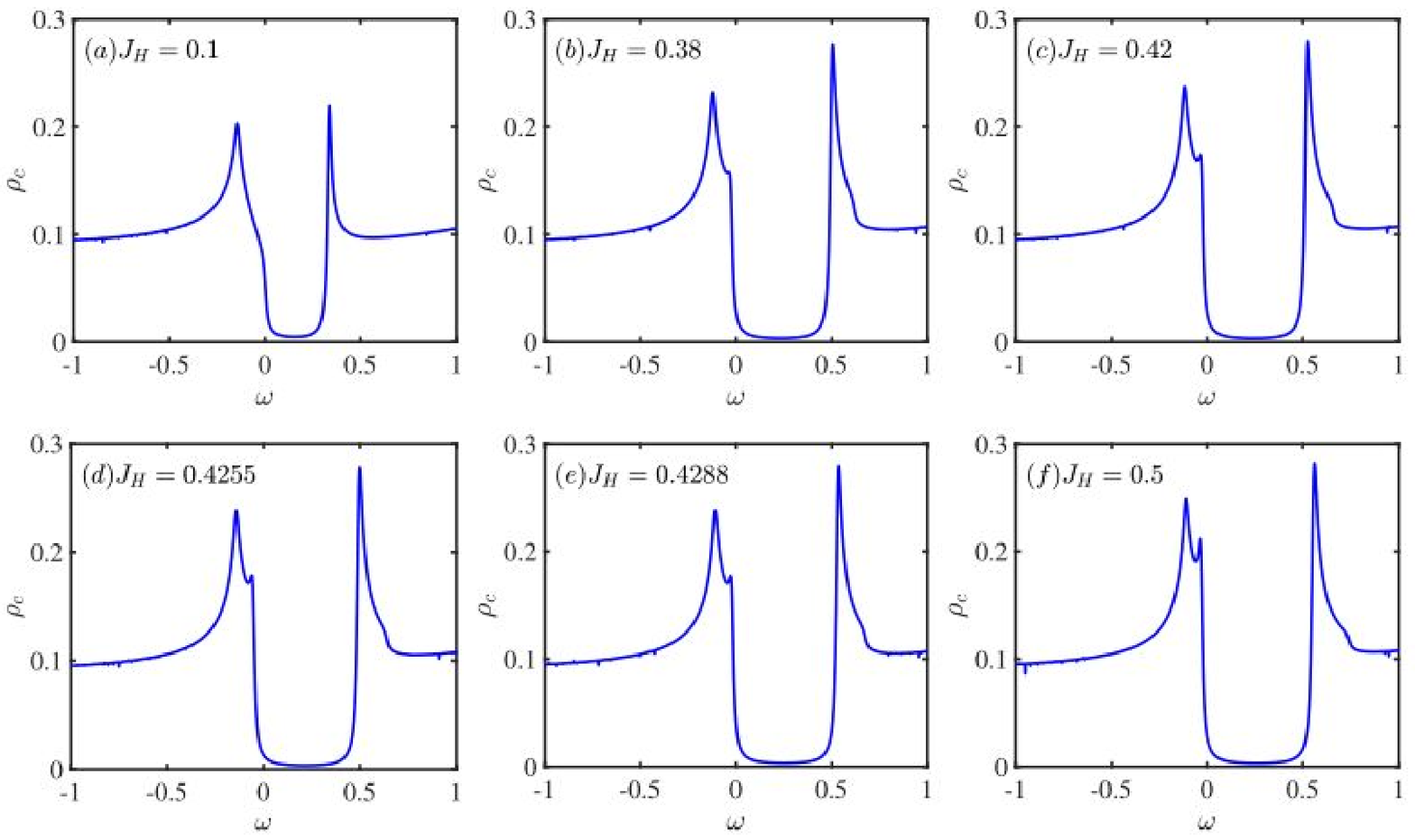}\\
\caption{\label{Fig3} The conduction electron DOS $\rho_{c}$ for the KHM versus the Heisenberg interaction $J_{H}$ for NN electron hopping but with NNN hole hopping, where $t=1, t_{1}/t=0.3, n_{c}=0.9, J_{K}=2.5$. It is shown that the conduction electron DOS is changed after Lifshitz transition.}
\end{figure}
\begin{figure}[htb]
\centering
\includegraphics[width = 0.8\columnwidth]{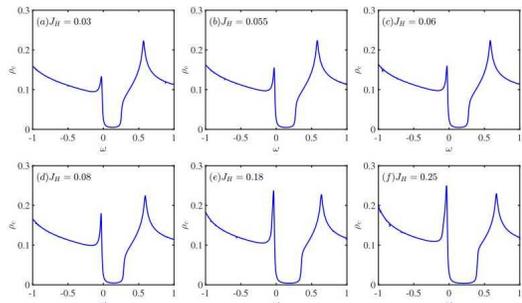}\\
\caption{\label{Fig4} The DOS of the conduction electron $\rho_{c}$ for the KHM versus the Heisenberg interaction $J_{H}$ for NN hole hopping but with NNN electron hopping, where $t=-1, t_{1}/t=0.3, n_{c}=0.9, J_{K}=2.5$. It is shown that the conduction electron DOS is changed after Lifshitz transition.}
\end{figure}

The DOS of the conduction electron is shown in Figs.~\ref{Fig3} - \ref{Fig4}. The Heisenberg interaction $J_{H}$ has an effect on the DOS of the conduction electron $\rho_{c}$, the larger $J_{H}$ induces the larger gap. Thus, the DOS is changed after Lifshitz transition. In Fig.~\ref{Fig3} (a)-(b), the DOS has a gap and two peaks. At $J_{H}=0.4255$,  it develops a new small peak as sown in Fig.~\ref{Fig3} (c), where the conduction electron DOS has one gap and three peaks when $J_{H}>0.4255$ as shown in Fig.~\ref{Fig3} (c)-(f). In Fig.~\ref{Fig4}, the DOS always has the gap and two peaks, but the left peak is lower than the right peak as shown in Fig.~\ref{Fig4} (a)-(d), while becomes higher than the right peak as shown in Fig.~\ref{Fig4} (e)-(f). Both peaks are increasing versus the Heisenberg coupling $J_{H}$, and the peak at $\omega<0$ that arises from the van Hove singularity of the large (hybridized) FS.\cite{Figgins2010}

Therefore, under the MF method,\cite{Liu2012,PhysRevB.83.033102} when Heisenberg superexchange $J_{H}$ increases, the presence of the short-range antiferromagnetic correlation gradually changes the electronic structure, and leads to the mentioned two kinds of Lifshitz transition, which is similar to Ref. \cite{PhysRevB.83.033102}. However, our work finds that the continuous transition around the Lifshitz transition, which is different from the square lattice, i.e. it has extra first-order transition.\cite{PhysRevB.83.033102}

With the FS topology of the quasiparticles changed, the area of FS varies at some critical values. To get more insight into the Lifshitz transition, it is helpful to use an effective low-energy theory to grasp the basic physical feature. Since the Lifshitz transition is mainly a single particle problem, one may use the following simple action
\begin{eqnarray}
S=\int d\tau d^{2}x\left[\psi^{\dag}\left(\partial_{\tau}-\frac{\nabla^{2}}{2M}-r\right)\psi\right],
\end{eqnarray}
where $M$, $r$ denote the effective mass and the effective chemical potential, respectively. The fermionic field $\psi$ represents the fermions whose FS will vanish (appear) when $r<0$ ($r>0$). Since the most radical effect of the Lifshitz transition is just such a/an disappearance/appearance of FS due to some parameters like $r$ here, we may expect this action captures the nature of this transition. When $r<0$, all fermions are gapped and no FS is observed while there exists a notable FS if $r>0$ is satisfied. At the transition point where $r=0$, the FS vanishes to a point and the corresponding local DOS is a constant. The specific heat at the transition point is $C_{v}\sim T$, which is undistinguished with the usual Fermi liquid's result.

Before ending this section, we note that the change of the FS topology, i.e. the Lifshitz transition, has a direct experimental implication. The Hall coefficient will change its sign when the electronic FS transforms into the hole-type one or some parts of FS disappear. Besides this, one can use the quantum oscillation to measure the effective mass of the quasi-particle as the signal of the Lifshitz transitions discussed here.

\section{The Differential conductance}\label{sec3}
The STM spectrum is one of the indispensable tools in the study of correlated quantum matter, especially for several quantum critical heavy-electron compounds, which is a real-space probe that measures a local conductance.\cite{RevModPhys.75.473,RevModPhys.79.353,kirchner2018arpes} In the linear-response regime, the current-voltage characteristics is related to the local DOS of the material.\cite{PhysRevB.31.805} There are also many STM experiments on the heavy-fermion compounds like YbRh$_{2}$Si$_{2}$ and CeCoIn$_{5}$.\cite{PhysRevX.5.011028,paschen2004hall,aynajian2012visualizing,allan2013imaging,zhou2013visualizing,JPSJ.81.011002} Those results coincide with angle-resolved photoemission spectroscopy to understand the physics of quantum critical point in heavy-fermion compounds.\cite{kirchner2018arpes}

Here, we follow Ref.~\cite{Figgins2010} to get the differential conductance $\frac{dI(V)}{dV}$ on the triangular lattice by
\begin{equation}
\frac{dI(V)}{dV}=t^{2}_{c}N_{cc}(V)+t^{2}_{f}N_{cc}(V)+2t_{c}t_{f}N_{cf}(V),
\end{equation}
where $t_{c}$ is the tunneling amplitude of the conduction electron and $t_{f}$ is for the $f$-electron. $N_{cc}=-\frac{1}{\pi}\text{Im}G_{cc}$ is the DOS of conduction electron while $N_{ff}=-\frac{1}{\pi}\text{Im}G_{ff}$ is for $f$-electron and their mixture is $N_{cf}=-\frac{1}{\pi}\text{Im}G_{cf}$. $G_{cc}(\tau)=-\langle T_{\tau}c_{k\sigma}(\tau)c^{\dag}_{k\sigma}(0)\rangle$ is the Green's function of conduction electron, the $G_{ff}(\tau)=-\langle T_{\tau}f_{k\sigma}(\tau)f^{\dag}_{k\sigma}(0)\rangle$ is for $f$-electron and $G_{cf}(\tau)=-\langle T_{\tau}c_{k\sigma}(\tau)f^{\dag}_{k\sigma}(0)\rangle$ describes the many-body effects arising from the hybridization of the conduction band with the $f$-electron level.

\begin{figure}[htb]
\centering
\includegraphics[width = 0.8\columnwidth]{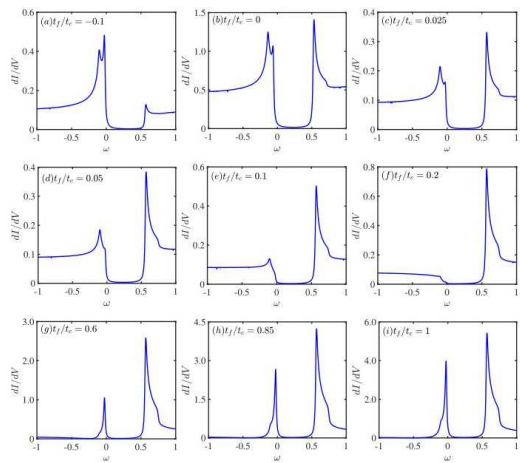}\\
\caption{\label{Fig5}  The STM spectra of the KHM versus the ratio of amplitudes $t_{f}/t_{c}$ for NN electron hopping but with NNN hole hopping, where $t=1, t_{1}/t=0.3, n_{c}=0.9, J_{K}=2.5$. The ratio of amplitudes $t_{f}/t_{c}$ influences the line-shape of the $dI/dV$.  }
\end{figure}

\begin{figure}[htb]
\centering
\includegraphics[width = 0.8\columnwidth]{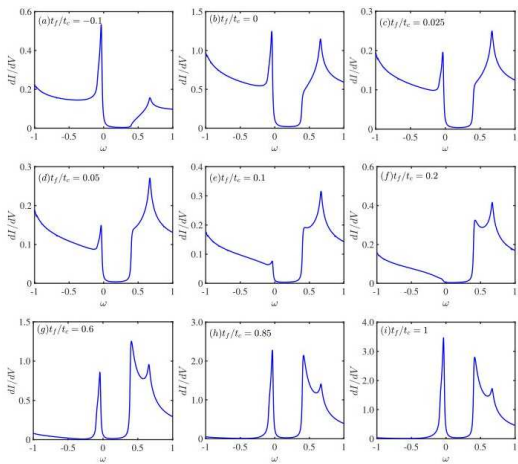}\\
\caption{\label{Fig6} The STM spectra of the KHM versus the ratio of amplitudes $t_{f}/t_{c}$ for NN hole hopping but with NNN electron hopping, where $t=-1, t_{1}/t=0.3, n_{c}=0.9, J_{K}=2.5$. The ratio of amplitudes $t_{f}/t_{c}$ influences the line-shape of the $dI/dV$.}
\end{figure}

To calculate DOS, it is helpful to introduce fermionic quasiparticle $A_{k\sigma}$ and $B_{k\sigma}$ with the following transformation
\begin{eqnarray}
&&f_{k\sigma}=v_{k}A_{k\sigma}+u_{k}B_{k\sigma}\label{eq10},\\
&&c_{k\sigma}=u_{k}A_{k\sigma}-v_{k}B_{k\sigma}\label{eq11},
\end{eqnarray}
where $u^{2}_{k}=\frac{1}{2}+\frac{-\chi_{k}+(\varepsilon_{k}-\mu)}{2\sqrt{(\varepsilon_{k}-\mu-\chi_{k})^{2}+(J_{K}V)^{2}}}$,  $v^{2}_{k}=1-\mu_{k}^{2}$, and $u_{k}v_{k}=\frac{J_{K}V}{2\sqrt{(\varepsilon_{k}-\mu-\chi_{k})^{2}+(J_{K}V)^{2}}}$.
The energy spectrum are given by Eq.~(\ref{eq3}). Fig.~\ref{Fig5} shows the shape-lines of the differential conductance $dI/dV$ in NN electron hopping with NNN hole hopping, and Fig.~\ref{Fig6} shows the case of NN hole hopping with NNN electron hopping. With increasing the ratio of amplitudes $t_{f}/t_{c}$, the hybridization between conduction electron band and $f$-electron band is different.

When increasing $t_{f}/t_{c}$, the line-shape changes quickly. Fig.~\ref{Fig5} (a) - (c) have three peaks and (h) - (i) have two peaks. There emerges a peak when $\omega>0$, and the peak is increasing versus the $t_{f}/t_{c}$, which is the precursor of the emerging f-electron band.\cite{Figgins2010} Fig.~\ref{Fig6} (a) - (d) have two peaks and (h) - (i) exist three peaks. Fig.~\ref{Fig5} (e) becomes two peaks while Fig.~\ref{Fig6} (e) begins to have three peaks. In the subplots (f) of Figs.~\ref{Fig5} - \ref{Fig6}, the left resonance peak nearly vanishes, which means the suppression of the differential conductance around the Fermi energy.\cite{Figgins2010}

\begin{figure}[htb]
\centering
\includegraphics[width = 0.8\columnwidth]{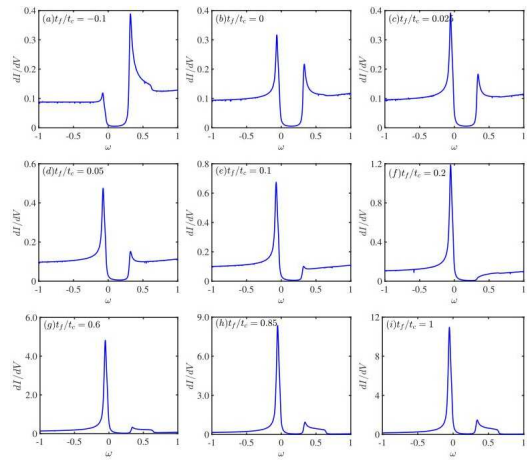}\\
\caption{\label{Fig7}  The STM spectra of the KHM versus the ratio of amplitudes $t_{f}/t_{c}$ for NN electron hopping but with NNN hole hopping, where $t=1, t_{1}/t=0.3, n_{c}=0.9, J_{K}=2$. The ratio of amplitudes $t_{f}/t_{c}$ influences the shape-line of the $dI/dV$.  }
\end{figure}

\begin{figure}[htb]
\centering
\includegraphics[width = 0.8\columnwidth]{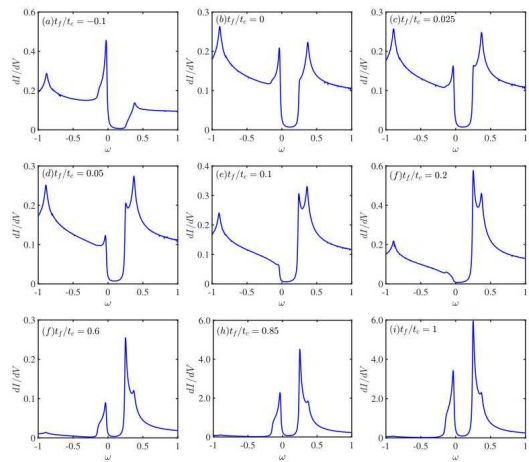}
\caption{\label{Fig8} The STM spectra of the KHM model versus the ratio of amplitudes $t_{f}/t_{c}$ for NN hole hopping but with NNN electron hopping, where $t=-1, t_{1}/t=0.3, n_{c}=0.9, J_{K}=2$. The ratio of amplitudes $t_{f}/t_{c}$ influences the shape-line of the $dI/dV$.  }
\end{figure}

We also give the STM spectra of the different Kondo coupling $J_{K}$ as shown in Figs.~\ref{Fig7} - \ref{Fig8}. Compared with Figs.~\ref{Fig5} - \ref{Fig6}, the line-shape varies versus $t_{f}/t_{c}$. There also exist the suppression of the differential conductance around the Fermi energy as shown in the subplots (e) of Figs.~\ref{Fig7}~-~\ref{Fig8}. The subplots (b) of Figs.~\ref{Fig5} - \ref{Fig8} are the DOS of the conduction electron.

Among those figures, the gap is increasing versus the Heisenberg coupling $J_{H}$ and the Kondo coupling $J_{K}$. STM line-shape of the differential conductance $dI/dV$ is mainly influenced by the ratio of $t_{f}/t_{c}$, the larger $t_{f}/t_{c}$ induces larger peak. The line-shape of the differential conductance $dI/dV$ emerges much f-electron information versus the ratio of $t_{f}/t_{c}$. We also find that those spectra are qualitatively similar with CeCoIn$_{5}$.\cite{aynajian2012visualizing} These results show that the existence of two resonance peaks structure in differential conductance as Refs.~\cite{Figgins2010,Maltseva2009}, which gives the insight to the heavy-fermion compounds by STM to examine the correlated electrons with high energy and spatial resolutions.\cite{Aynajian2012}

%With the gap becoming larger versus the $J_{H}$, the short-range antiferromagnetic correlation causes the quasiparticles effective masses to increase. When $J_{K}=2$ and $n_{c}=0.85$ (the figures g), h) and i) in the Fig.\ref{Fig6}), the shape-lines are the similar to the figures a), b) and c) in the Fig.\ref{Fig5}, while the gap between hybridized bands is smaller than before due to smallness of conduction electron density.


\section{Conclusion and perspective}\label{sec5}
In summary, we have investigated the KHM on triangular lattice with the fermionic large-N mean-field theory at the case of the NN electron hopping with NNN hole hopping and the case of the NN hole hopping and NNN electron hopping. At the heavy-fermion liquid state, the Heisenberg antiferromagnetic interaction ($J_H$) induces twice FS topology change, i.e. the Lifshitz transition, where goes through the continuous transition. In two cases, the conduction electron DOS is changed after Lifshitz transition, the gap is influenced by the Kondo coupling $J_{K}$ and the Heisenberg interaction $J_{H}$. The line-shape of the differential conductance $dI/dV$ shows that the existence of two resonance peaks structure in differential conductance as Refs.~\cite{Figgins2010,Maltseva2009}.The short-range antiferromagnetic correlation coupling $J_{H}$, the ratio of the amplitudes of the f-electron to the amplitude of the the conduction electron $t_{f}/t_{c}$, and the Kondo correlation $J_{K}$ influence the shape-line of the differential conductance $dI/dV$, which gives the insight to detect the heavy-fermion compounds STM spectra for examining the correlated electrons with high energy and spatial resolutions.\cite{Aynajian2012}

Owing to some triangular heavy-fermion compounds like YbAgGe\cite{PhysRevB.71.054408,PhysRevLett.110.176402,PhysRevB.69.014415,PhysRevLett.111.116401} and YbAl$_{3}$C$_{3}$\cite{PhysRevB.87.220406,PhysRevB.85.144416} have been found, we expect that our results may be confirmed by many FS measurements (Hall coefficient, de Haas-van Alphen measurements, angle-resolved photoemission spectroscopy, quasiparticle interference and STM spectrum experiments) in those compounds.

\section*{Acknowledgments}
This work was initially projected by Yu-Feng Wang, and most of it has been finished by authors listed here. The authors acknowledge useful result given by Wang. This research was supported in part by NSFC under Grant No. 11674139, No. 11704166, No. 11834005, the Fundamental Research Funds for the Central Universities, and PCSIRT (Grant No. IRT-16R35).

\end{document}